# Lateral Spin Injection in Germanium Nanowires

*En-Shao Liu, Junghyo Nah, Kamran M. Varahramyan, Emanuel Tutuc\**
Microelectronics Research Center, The University of Texas at Austin,
Austin, TX 78758 USA
\* To whom correspondence should be addressed. E-mail: etutuc@mer.utexas.edu

Electrical injection of spin-polarized electrons into a semiconductor, large spin diffusion length, and an integration friendly platform are desirable ingredients for spin-based devices. Here we demonstrate lateral spin injection and detection in germanium nanowires, by using ferromagnetic metal contacts and tunnel barriers for contact resistance engineering. Using data measured from over 80 samples, we map out the contact resistance window for which lateral spin transport is observed, manifestly showing the conductivity matching required for spin injection. Our analysis, based on the spin diffusion theory, indicates that the spin diffusion length is larger than 100 μm in germanium nanowires at 4.2 K.

Electrical spin injection in semiconductors is fundamental in enabling information processing using the electron spin degree of freedom.[1-4] Several device concepts,[5-12] all employing ferromagnetic contacts (FMs) to semiconductors (SCs), have been advanced for logic operations and use the spin as the information-carrying degree of freedom. Efficient injection of spin-polarized electrons from FMs into SCs, a prerequisite for spin information processing, is typically suppressed by the mismatched conductivities between FMs and SCs.[13,14] Indeed, even in the absence of spin relaxation in the semiconductor, owing to the $10^3$-$10^4$ -fold higher metal conductivity the spin polarization of electrons flowing across a typical ferromagnetic metal-semiconductor interface can hardly exceed 0.1% (ref. 13).

There has been significant progress recently to achieve spin injection in semiconductors, either by inserting a spin-dependent tunnel barrier at the FM/SC interface[15-18] or by using hot-electron injection.[19] Group IV semiconductors, such as diamond, graphene, silicon, have attracted significant interest as a spintronic materials owing to several attributes.

The inversion symmetric crystal structure of these semiconductors reduces the spin-orbit induced spin splitting and translates into large spin diffusion length.[20] For carbon and silicon, the low count of isotopes with non-zero nuclear spins suppresses the hyperfine interaction-induced spin relaxation. Lastly, their compatibility with microelectronics technology makes them attractive for potential applications.

In this study we demonstrate efficient electrical spin injection and detection in germanium nanowires (Ge NWs), a finding which highlights germanium as a potential platform for spin-based devices. Germanium has an inversion symmetric crystal structure, which results in a weak spin-orbit coupling and consequently slow spin relaxation and large spin diffusion length. The additional confinement in nanowires may further suppress spin relaxation.[21,22] Indeed, for our lateral n-type Ge NW spin injection devices we extract a spin diffusion length larger than 100 μm at 4.2K.

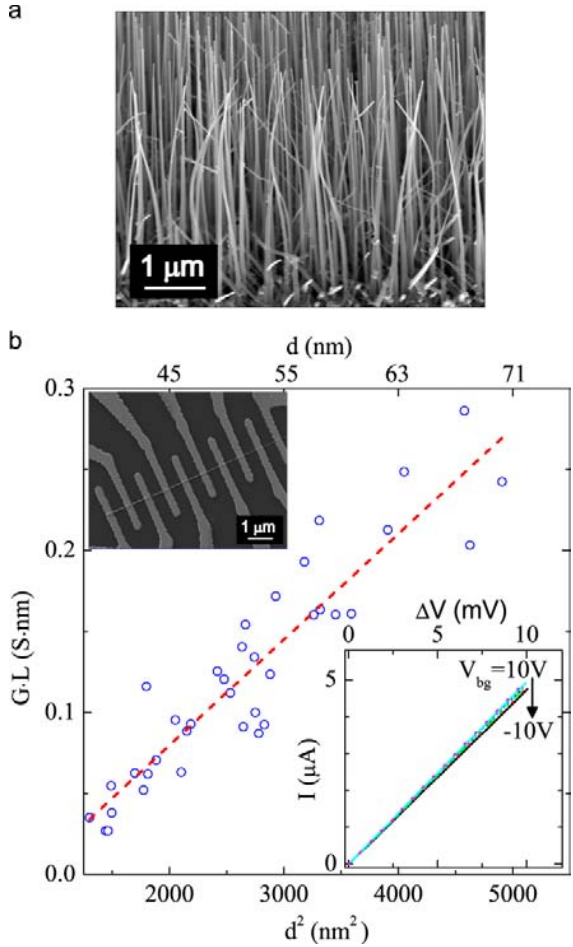

FIGURE 1. Structural and electrical properties of phosphorous-doped Ge NWs. (a) Scanning electron micrograph (SEM) of the phosphorous-doped Ge NWs epitaxially grown on a Si substrate. (b) $G·L$ product $vs. d^2$, measured at $T = 4.2$ K, for the Ge NWs examined in this study. The linear dependence on $d^2$ implies that the doping concentration along the NW is nearly constant for $d \geq 40$ nm. Upper inset: SEM of a multi-terminal Ge NW FET. Lower inset: four-point $I$-$V$ characteristics of a Ge NW FET, measured for $V_{bg} = 10$V to $-10$V.

Equally noteworthy, the smaller Ge band-gap allows for contact resistance engineering to values that allow for spin injection more efficient than in Si. As we show below, the Co-Ge specific contact resistance can be as low as $\sim 1 \times 10^{-8}$ $\Omega \cdot cm^2$ for NW resistivities of $\sim 2 \times 10^{-3}$ $\Omega \cdot cm$; within the framework of spin diffusion theory these values would allow for spin injection as long as the spin diffusion length is larger than $\sim 50$ nm.[14,23]

The Ge NWs are epitaxially grown on Si (111) substrates in an ultra-high vacuum chemical vapor deposition chamber via the gold-seeded, vapor-liquid-solid mechanism.[24] The phosphorus-doped Ge NWs are grown at a total pressure of 5 Torr and a wafer temperature of 400 °C using 100 sccm of $GeH_4$ (20% dilution in helium) and 10 sccm of $PH_3$ (100 ppm dilution). A 90 min growth yields $\sim 12$ μm-long Ge NWs, with base diameters between 80 and 90 nm and tip diameters between 20 and 30 nm (Fig. 1a). The NWs are subsequently harvested onto a 25 nm-thick $SiO_2$ film, thermally grown on a heavily doped p-type Si substrate, which serves as the back-gate for our devices. In order to characterize the electrical and doping properties, the NWs are fabricated into multi-terminal NW field-effect transistors (FETs) with Co contacts, using e-beam lithography and liftoff (Fig. 1b, upper inset). A 10 nm-thick gold film is deposited on top of Co to prevent post-processing oxidation.

We use four-point measurements to determine the NW conductance ($G$) and metal/NW contact resistance ($R_c$). An example of four-point current ($I$) vs. voltage ($V$) data is shown in the lower inset of Fig. 1b. Figure 1b shows the NW conductance and channel length ($L$) product, measured at a temperature $T = 4.2$ K, plotted versus the NW diameter ($d$) square. The linear dependence of these two quantities indicates that the doping density is constant for the diameter range investigated. The doping concentration ($n$) can be estimated from where $(G \cdot L)|_{V_{bg}=0} = \pi e \mu n d^2 / 4$, $\mu$ is the electron mobility, and $V_{bg}$ is the back-gate bias. The mobility values are extracted from the measured $G$ dependence on the applied back-gate bias ($V_{bg}$), using $\mu = C_{ox}^{-1} \cdot d(G \cdot L)/dV_{bg}$; here $C_{ox}$ is the back-gate to NW capacitance per unit length.

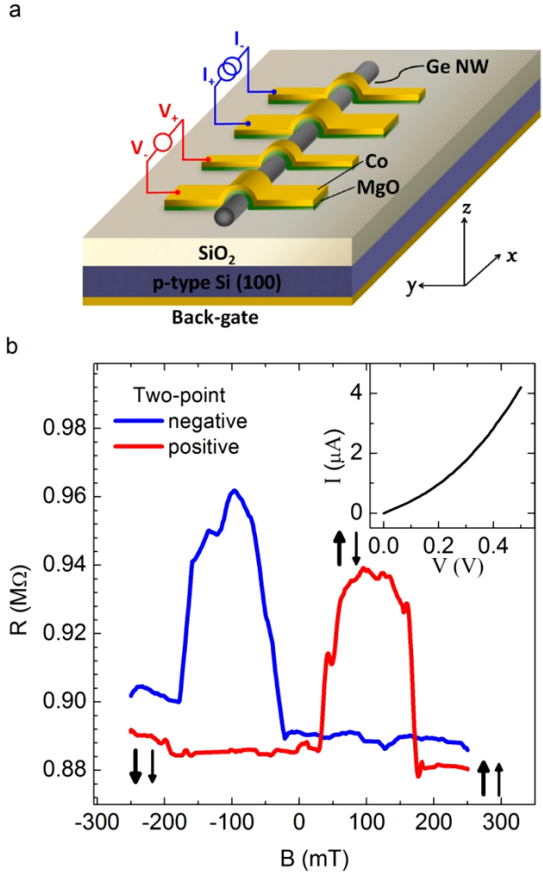

FIGURE 2. Schematic of Ge NW spin injection devices and the MR data. (a) Schematic representation of a Ge NW spin injection device with MgO tunnel barrier and Co electrodes. In a nonlocal measurement the current is injected from $I_+$ to $I_-$, and the voltage difference $(V_+ - V_-)$ is measured between the two contacts outside of the current path. (b) $R$ vs. in-plane $B$-field measured in the two-point configuration. The red (blue) trace corresponds to the positive (negative) sweep direction. The arrows indicate the magnetization directions of the contacts. Inset: two-point $I$-$V$ characteristic of the investigated Ge NW device.

The $C_{ox}$ values, calculated using self-consistent numerical simulations (Sentaurus), range between 74 and 91 aF/µm, for $d$ values between 41 and 70 nm. The extracted mobility in our NWs is $70\pm20$ cm$^2$/Vs, and the doping concentration is $5\pm2\times10^{19}$ cm$^{-3}$. Noteworthy, the contact resistances ($R_c$) between Co and the Ge NWs are very low, which allows for interface resistance engineering to overcome the conductivity mismatch. Indeed, the $R_c$ values for our Ge NWs with Co contacts are $300\pm100$ Ω, corresponding to specific contact resistance ($\rho_c$) of $1.8\pm1.6\times10^{-8}$ Ω·cm$^2$ (ref. 25). The fabrication of spin injection devices is similar to the above process; the key difference is that, without breaking vacuum, a thin layer of MgO is deposited by e-beam evaporation on the NW *prior to* Co deposition (Fig. 2a).

We now turn to the magnetoresistance (MR) and the spin valve effect in Ge NWs. The measurements are performed using low frequency lock-in techniques, at $T$ = 4.2K. As shown schematically in Fig. 2a, the multi-finger structure allows us to investigate the MR in the two-point (*local*) configuration, and in the *nonlocal* configuration.[26] In the latter (Fig. 2a) the current flows between the $I_+$ and $I_-$ contacts, and a voltage difference is measured between $V_+$ and $V_-$, outside of the current path. The nonlocal setup detects the spatial dependence of the spin-dependent chemical potentials created by the accumulation, diffusion, and relaxation of electron spins,[27] and excludes other effects that might lead to the same signature as the spin valve (e.g. the magneto-Coulomb effect[28]). The Co electrodes are designed to have different widths, with the narrower contacts at 350 nm and the wider ones at 600 nm. This ensures that as a function of a parallel magnetic field ($B$), the electrode magnetization will reverse at different $B$-fields: the larger the width, the lower the exchange energy barrier, and hence the smaller the coercive field.[29] Consequently, the device MR can be probed at different electrode magnetization configurations.

Figure 2b data show an example of the two-point resistance ($R$) of a lateral Ge NW device with a 10 Å-thick MgO layer, as a function of the $B$-field applied parallel to the electrodes. Figure 2b inset data show the $I$ vs.$V$ measured for the same device; owing to the MgO barrier the Co/MgO/NW contact has a higher contact resistance and shows a slight non-linearity in $I$-$V$ curve. The MgO tunnel barrier thickness, hence $R_c$, proves crucial to achieve spin injection. With the thickness being varied from 0 to 20 Å, we can tune $R_c$ from 170 Ω to more than 50 MΩ. As we discuss below, the spin valve effect can only be observed for a narrow window of $R_c$ values.

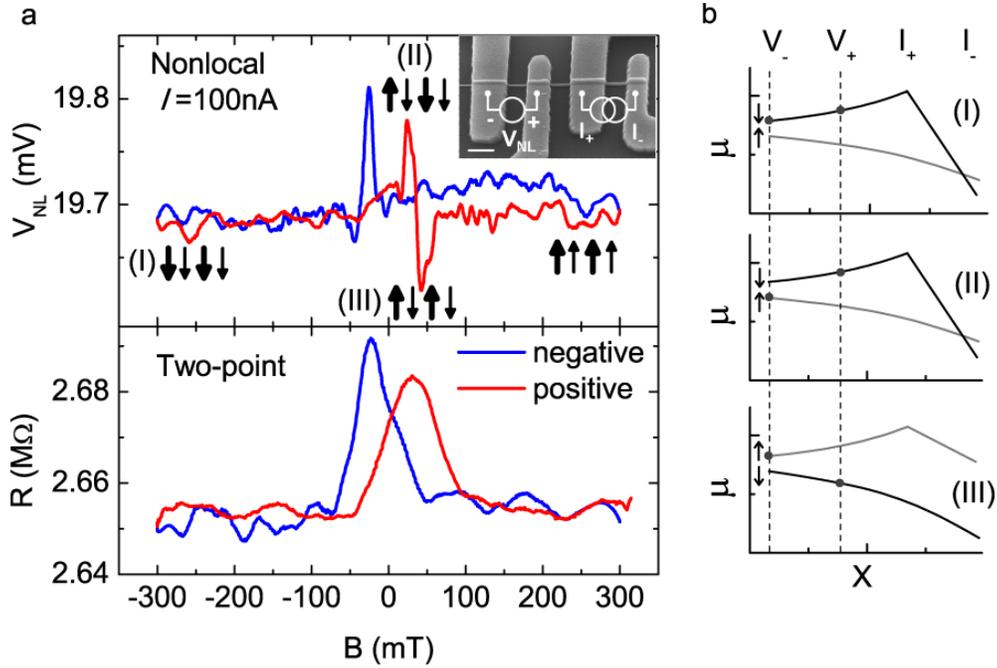

FIGURE 3. Spin signal in nonlocal and two-point configuration, and schematics of spatial-dependent $\mu_\uparrow$ and $\mu_\downarrow$ at different magnetization configurations. (a) Top panel: Nonlocal voltage ($V_{NL}$) as a function of the in-plane $B$-field for positive and negative sweep direction. At large negative $B$, all four electrodes' magnetization directions are parallel. As $B$ is swept toward the positive direction, the signal jumps to a maximum when the magnetization direction of the $V_-$ electrode switches and becomes antiparallel to other three contacts. The $I_-$ contact switches magnetization as $B$ is further increased, and the signal drops. At larger $B$ all contacts magnetizations are parallel and the signal returns to background value. Inset: SEM of the Ge NW device, and the nonlocal measurement configuration. Bottom panel: Two-point MR data measured between the two contacts used as current leads in the nonlocal measurement. The resistance peaks when the two contacts have antiparallel magnetizations, and occurs at the same $B$-field where the transitions happen in the nonlocal traces. (b) Schematics of the $\mu_\uparrow$ and $\mu_\downarrow$ along the NW; the dots indicate the spin orientation probed by the voltage contacts.

For the positive sweep (red curve) of Fig. 2b, at $B = -250$ mT, both electrode magnetizations are aligned with the $B$-field. As $B$ is increased to $+50$ mT, the magnetization directions of the two electrodes become antiparallel as the wider electrode changes polarization, and the resistance increases by 60 kΩ. The resistance stays constant until $B$ reaches $+160$ mT, at which the magnetization of the narrow electrode reverses. Sweeping the $B$-field further, the electrodes' magnetizations become parallel again and the resistance falls back to the initial value. The reverse sweep generates a symmetric trace.

In order to verify that the observed spin valve-like signal in the two-point configuration stems from spin injection, we performed MR measurements in the nonlocal configuration.

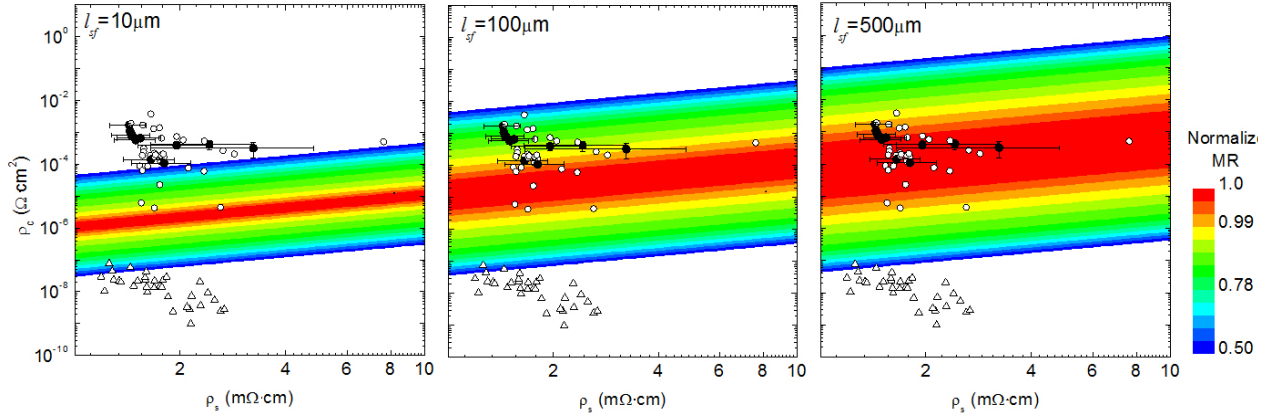

FIGURE 4. Specific contact resistance ($\rho_c$) vs. resistivity ($\rho_s$) data for Ge NWs with Co contacts. The different symbols represent devices with (circles) or without (triangles) MgO tunnel barriers. The closed symbols represent devices that exhibit spin valve effect. The MR contour plot is calculated[14] using $l_{sf}$ = 10, 100, and 500 μm, from left to right, respectively. The maximum MR contour (red corridor) overlaps best with the devices showing spin injection for $l_{sf}$ = 500 μm; partial overlap is obtained as long as $l_{sf}$ is assumed to be larger than 100 μm. We note that some devices in the red band did not exhibit spin valve effect, a finding we attribute to variability associated with e-beam evaporation of MgO, namely lack of crystallinity or a well defined crystal direction of the Co/MgO/Ge NW stack.

Figure 3a inset shows a scanning electron micrograph (SEM) of the device and the contact configuration used in the nonlocal MR measurement; the device contains a 15 Å-thick MgO tunnel barrier. The top panel of Fig. 3a shows the nonlocal voltage difference ($V_{NL} \equiv V_+ - V_-$) vs. B. The measured signal can be explained by examining the correspondence between the magnetizations of the contacts and the spin-up ($\mu_\uparrow$) and spin-down ($\mu_\downarrow$) chemical potentials. At B = -300 mT, all four electrodes are magnetized toward the negative direction, as indicated by configuration (I). The spin-polarized electrons injected into the NW create spatial-dependent $\mu_\uparrow$ and $\mu_\downarrow$ along the NW axis, as shown in the top panel of Fig. 3b. The constant background signal in Fig. 3a is typically observed in nonlocal measurements, independent of the spin valve effect, and decreases with increasing distance between the current and voltage probes. The background may originate from the small leakage currents between large contact pads and the substrate, or may be related to the non-uniform electron injection at the contacts.[30] As B is ramped to +23 mT [configuration (II) of Fig. 3a, and middle panel of Fig. 3b], $V_-$ reverses and detects $\mu_\uparrow$ while $V_+$ still senses $\mu_\downarrow$. This translates into a 90 μV increase in $V_{NL}$. At B = 41 mT, the $I_+$ electrode switches and is now injecting spin-up electrons into the NW [configuration (III) in Fig. 3a, and the bottom panel of Fig. 3b]. While $V_+$ and $V_-$ are still sensitive to $\mu_\downarrow$ and $\mu_\uparrow$, respectively, the voltage difference is now of the same magnitude but opposite sign to that of the previous stage, which translates into the 70 μV drop below the background level in Fig. 3a. At even larger B-field all the electrodes' magnetizations are aligned, and the signal now represents the spatial dependence of $\mu_\uparrow$. Reverse sweeps show similar behaviour. Also shown in the bottom panel of Fig. 3a is the two-point MR data measured between electrodes $I_+$ and $I_-$, which behaves similarly to that of Fig. 2b: the resistance initially stays constant while the field is slowly being swept toward the opposite direction, jumps to a larger value when the wider electrode flips its magnetization, and drops to the initial value after both electrodes are again aligned with the field. The transitions in two-point and nonlocal data occur at the same *B*-field, which strongly suggests that the spin valve effect observed in two-point MR originates from spin injection and accumulation in Ge NWs.

A key parameter to describe the spin transport is the electron's spin diffusion length ($l_{sf}$), namely the distance that an electron can travel before losing its spin orientation. The reported $l_{sf}$ values in other semiconductors are, 1.8 μm in GaAs (ref. 14), and 2 μm in graphene[18] at low (< 10 K) temperatures. For Si, coherent spin transport over 10 μm was demonstrated using hot electron injection at 85K (ref. 19), and recently a $l_{sf}$ of 0.2 μm at room temperature has been reported.[15] The $l_{sf}$ value in a semiconductor is related to the two-point MR (≡ Δ*R*/*R*$_P$) according to Fert and Jaffrès[14]:

$$\Delta R = \frac{2(\beta r_{Co} + \gamma \rho_c)^2}{\rho_c + r_{Co} + \left[1 + \left(\frac{\rho_c}{r_{NW}}\right)^2\right] \cdot \frac{L \cdot r_{NW}}{2 l_{sf}}}$$

$$R_p = 2(1-\beta^2)r_{Co} + \frac{L \cdot r_{NW}}{l_{sf}} + 2(1-\gamma^2)\rho_c + 2\frac{(\beta-\gamma)^2 r_{Co}\rho_c + r_{NW}(\beta^2 r_{Co}+\gamma^2\rho_c)\tanh\left(\frac{L}{2l_{sf}}\right)}{r_{Co}+\rho_c+r_{NW}\tanh\left(\frac{L}{2l_{sf}}\right)}$$

where Δ*R* is the resistance difference between the antiparallel ($R_{AP}$) and parallel ($R_P$) configurations of the electrodes' magnetizations, β and γ are the bulk spin asymmetry coefficient in a Co electrode and spin-dependent tunnelling coefficient of the Co/MgO/Ge NW contact, $r_{NW}$ and $r_{Co}$ are the product of $l_{sf}$ and resistivity of the Ge NWs and Co, respectively. The parameters in these equations are either known, such as β and $r_{Co}$ (ref. 14), or measured in these experiments, such as the Ge NW resistivity ($\rho_s$), and $\rho_c$. However, the spin-dependent tunnelling coefficient remains unclear for the Co/MgO/Ge NW tunnel contact used here. Moreover, owing to the absence of a well defined crystal direction at the Co/MgO/GeNW contact, as well as the e-beam evaporated MgO, we expect the γ values to be device dependent.

In order to estimate the $l_{sf}$ in Ge NWs we examined more than eighty devices spanning over six orders of magnitude in $\rho_c$, and manifestly mapped out the optimum conditions for spin injection. Figure 4 shows $\rho_c$ vs. $\rho_s$ for all devices examined in this study; the closed (open) symbols represent devices in which spin valve effect is present (absent). Figure 4 data show that spin injection is only observed in devices with $\rho_c$ between 10$^{-4}$ and 10$^{-3}$ Ω·cm$^2$, and are absent at higher or lower $\rho_c$. We then calculated the optimal range of $\rho_c$ and $\rho_s$ values for spin injection (red corridor in Fig. 4) using γ and $l_{sf}$ as fitting parameters; higher (lower) $l_{sf}$ values move this corridor upward (downward), while γ impacts mainly the MR value.[14] In order to overlap the calculated ($\rho_c$, $\rho_s$) corridor which allows for spin injection with the *measured* ($\rho_c$, $\rho_s$) window where spin valve effects are experimentally observed, the $l_{sf}$ values in the Ge NWs examined here have to be at least 100 μm. As shown in Fig. 4, the best overlap between theory and experiment is obtained for $l_{sf}$ = 500 μm. Though the $l_{sf}$ cannot be determined more accurately using this technique, and further studies, such as spin precession measurements are needed for more precise values, our data strongly suggests that the spin diffusion length in Ge NWs at 4.2 K is ~100 μm or larger.

In summary, we demonstrate electrical spin injection and detection in n-type Ge NWs, and mapped out the contact resistance window which allows for spin injection, manifestly showing the conductivity matching required for spin injection. By exploring a wide parameter space in contact resistivity, we show that the spin diffusion length in Ge NWs is larger than 100 μm. These findings highlight Ge NWs as a potential spintronic material.

ACKNOWLEDGEMENT

We thank S. Banerjee, A. Schul, N. Samarth, and J. Petta for discussions. This work is supported by NSF grants DMR-0819860 and DMR-0846573, and by NRI-SWAN.